\documentclass[aip,jcp,reprint,amsmath,amssymb,floatfix,citeautoscript,nofootinbib]{revtex4-2}
\usepackage{cancel}
\usepackage{amsmath}
\usepackage{physics}
\usepackage{graphicx}
\usepackage{amssymb}
\usepackage{amsthm}
\usepackage{bm}
\usepackage{dcolumn}
\newcolumntype{x}[1]{D{.}{.}{#1}}

\usepackage{ragged2e}
\usepackage{txfonts}
\allowdisplaybreaks

\usepackage{color}
\usepackage[usenames,dvipsnames]{xcolor}
\definecolor{myblue}{rgb}{0,0,1}
\usepackage[breaklinks=true,colorlinks=true,linkcolor=myblue,urlcolor=myblue,citecolor=myblue]{hyperref}

\newcommand{\vk}{{\bm{k}}}
\newcommand{\vG}{{\bm{G}}}

\begin{document}

\title{
Accurate thermochemistry of covalent and ionic solids from spin-component-scaled MP2
}

\author{Tamar Goldzak$^*$}
\altaffiliation{Present address: Faculty of Engineering, Bar-Ilan University, Ramat Gan, Israel} 
\affiliation{Department of Chemistry, Columbia University, New York, NY 10027 USA}
\author{Xiao Wang$^*$}
\altaffiliation{Present address: Department of Chemistry and Biochemistry, University of California, Santa Cruz, CA 95064 USA}
\affiliation{Center for Computational Quantum Physics, Flatiron Institute, New York, NY 10010 USA}
\author{Hong-Zhou Ye}
\affiliation{Department of Chemistry, Columbia University, New York, NY 10027 USA}
\author{Timothy C. Berkelbach}
\email{t.berkelbach@columbia.edu}
\affiliation{Department of Chemistry, Columbia University, New York, NY 10027 USA}
\affiliation{Center for Computational Quantum Physics, Flatiron Institute, New York, NY 10010 USA}

\begin{abstract}
We study the performance of spin-component-scaled second-order M\o ller-Plesset
perturbation theory (SCS-MP2) for the prediction of the lattice constant, bulk
modulus, and cohesive energy of 12 simple, three-dimensional, covalent and ionic
semiconductors and insulators.  We find that SCS-MP2 and the simpler scaled
opposite-spin MP2 (SOS-MP2) yield predictions that are significantly improved
over the already good performance of MP2.  Specifically, when compared to
experimental values with zero-point vibrational corrections, SCS-MP2 (SOS-MP2)
yields mean absolute errors of 0.015 (0.017)~\AA\ for the lattice constant, 3.8
(3.7)~GPa for the bulk modulus, and 0.06 (0.08)~eV for the cohesive energy,
which are smaller than those of leading density functionals by about a factor of
two or more.  We consider a reparameterization of the spin scaling parameters
and find that the optimal parameters for these solids are very similar to those
already in common use in molecular quantum chemistry, suggesting good
transferability and reliable future applications to surface chemistry on
insulators.
\end{abstract}

\maketitle

\def\thefootnote{*}\footnotetext{These authors contributed equally}\def\thefootnote{\arabic{footnote}}

\section{Introduction}

In recent years, wavefunction-based methods, popular in the quantum chemistry
community, have begun to be regularly applied to condensed-phase systems.  In
particular, coupled-cluster theories, such as CCSD or CCSD(T)---the so-called
gold standard of molecular quantum chemistry---yield promising
results~\cite{Hirata2004,Grueneis2011,Booth2013,Grueneis2015,McClain2017,Tsatsoulis2017,gruber2018,gruber2018a,zhang2019,Gao2020,Nusspickel2022}.
However, due to their high scaling with system size $N$, i.e., $O(N^6)$ for CCSD
and $O(N^7)$ for CCSD(T), their applicability to solids with complex unit cells
is limited, and obtaining results at the thermodynamic limit and the complete
basis set limit is challenging.

As a lower cost alternative to coupled-cluster theory, second-order 
M\o ller-Plesset perturbation theory (MP2) is widely used in molecular quantum
chemistry and has been increasingly applied to periodic
systems~\cite{Ayala2001,Hirata2004,Hirata2009,Marsman2009,Gruneis2010,Usvyat2010,Maschio2010,Katouda2010,Grueneis2011,DelBen2012,Goeltl2012,DelBen2013,McClain2017,Schaefer2017,Tsatsoulis2017,Bintrim2022}.
Over a decade ago, Gr\"uneis et al.~\cite{Gruneis2010} demonstrated that, for
simple covalent and ionic solids, ground-state structural and electronic
properties predicted by MP2 are quite good and better than those predicted by
DFT with the popular Perdew-Burke-Ernzerhof (PBE) exchange correlation
functional~\cite{perdew1996generalized}. However, it is natural to consider
improvements to MP2, the limitations of which are well known in molecular
quantum chemistry. 

At the same cost, spin-component-scaled (SCS) MP2~\cite{scs_mp2_grimme2003},
which semiempirically scales the same-spin and opposite-spin components of the
correlation energy, has been shown to significantly outperform MP2 for many
molecular
properties~\cite{scs_mp2_grimme2003,hyla2004comprehensive,Grimme2005,Antony2007,takatani2007performance,distasio2007optimized,kossmann2010correlated,grimme2012spin,Steinmetz2013}.
Scaled opposite-spin (SOS) MP2~\cite{sos_mp2_head_gordon}, which retains and
scales only the opposite-spin component of the correlation energy, shows
comparably good
performance~\cite{sos_mp2_head_gordon,Grimme2005,grimme2012spin,Steinmetz2013}
and can be performed with a reduced $O(N^4)$ scaling using density fitting and a
Laplace transform of the energy denominators. Both SCS- and SOS-MP2 commonly
deliver an accuracy that is better than CCSD and comparable to CCSD(T),
suggesting that they could be especially promising methods for complex,
insulating solids.  
 
Here, we test periodic SCS/SOS-MP2 for the calculation of the lattice constant,
bulk modulus, and cohesive energy of 12 three-dimensional, covalent and ionic
semiconductors and insulators, paying careful attention to the thermodynamic
limit and complete basis set limit.  The layout of this paper is as follows. In
Sec.~\ref{sec:methods}, we briefly review the formalism of SCS/SOS-MP2 and
provide details of our periodic MP2 calculations, basis sets, and density
fitting.  We also present the convergence of the HF energy and the components of
the MP2 correlation energy to the thermodynamic limit and complete basis set
limit, using carbon diamond as an example.  In Sec.~\ref{sec:results}, we
present the accuracy of the calculated properties for 12 solids, including a
scan over the spin-scaling parameters.  We conclude in
Sec.~\ref{sec:conclusion}.

\section{Methods}
\label{sec:methods}

The SCS-MP2 correlation energy can be split into two components,
\begin{equation}
\label{eq:ec}
E^{(2)} = 
    c_\mathrm{os} E_\mathrm{os}^{(2)}
    + c_\mathrm{ss} E_\mathrm{ss}^{(2)}
\end{equation}
where the spin-scaling parameters $c_\mathrm{os}=c_\mathrm{ss}=1$ for
traditional MP2.  On the basis of molecular data, the conventional SCS-MP2
parameters are 
$(c_\mathrm{os}, c_\mathrm{ss}) = (1.2, 0.33)$~\cite{scs_mp2_grimme2003} 
and the conventional SOS-MP2 parameter is
$c_\mathrm{os} = 1.3$ (with $c_\mathrm{ss}=0$)~\cite{sos_mp2_head_gordon},
although reoptimization has been explored in a variety of
contexts~\cite{distasio2007optimized,Rigby2014,Tan2017}.

With periodic boundary conditions and $N_k$ crystal momenta $\vk$ sampled from
the Brillouin zone, the spin components of the correlation energy are
\begin{subequations}
\begin{align}
E_\mathrm{os}^{(2)} &= \frac{1}{N_k^3} \sum_{\vk_i\vk_a\vk_j\vk_b}^\prime \sum_{iajb}
    T_{i \vk_i,j \vk_j}^{a\vk_a, b\vk_b}
    (i \vk_i a \vk_a | j \vk_j b \vk_b), \\
\begin{split}
E_\mathrm{ss}^{(2)} &= \frac{1}{N_k^3} \sum_{\vk_i\vk_a\vk_j\vk_b}^\prime 
    \sum_{iajb} 
    \left[ T_{i \vk_i,j \vk_j}^{a \vk_a, b\vk_b} 
        - T_{i \vk_i,j \vk_j}^{b \vk_b, a\vk_a} \right] \\
&\hspace{8em} \times
    (i\vk_i a\vk_a | j\vk_j b\vk_b),
\end{split}
\end{align}
\end{subequations}
where
\begin{equation}
 T_{i \vk_i,j\vk_j}^{a\vk_a,b\vk_b}
    = \frac{(i \vk_i a \vk_a|j\vk_j b\vk_b)^*}
        {\varepsilon_{i\vk_i} + \varepsilon_{j\vk_j} - \varepsilon_{a\vk_a} - \varepsilon_{b\vk_b}}.
\end{equation}
Electron repulsion integrals are given in Mulliken $(11|22)$ notation and, as
usual, $i,j$ refer to occupied orbitals and $a,b$ refer to unoccupied orbitals
in the periodic Hartree-Fock (HF) determinant, which we assume to be
spin-restricted.  The primed summation indicates conservation of crystal
momentum, $\vk_i+\vk_j-\vk_a-\vk_b=\vG$, where $\vG$ is a reciprocal lattice
vector. 

All calculations were performed with PySCF~\cite{Sun2017,Sun2020}.  The
Brillouin zone was sampled with a uniform Monkhorst-Pack mesh of $N_k$
$k$-points that includes the $\Gamma$ point.  Core electrons were replaced with
correlation-consistent effective-core potentials (ccECPs) and we used their
corresponding correlation-consistent cc-pV$X$Z basis
sets~\cite{bennett2017,bennett2018} ($X$Z).  For Li and Mg, we used the
large-core pseudopotentials ([He] core and [Ne] core, respectively).  Gaussian
density fitting with an even-tempered auxiliary basis set was used for
evaluating the electron repulsion integrals~\cite{Sun2017a}.

\begin{figure}[t]
\centering
\includegraphics[scale=0.9]{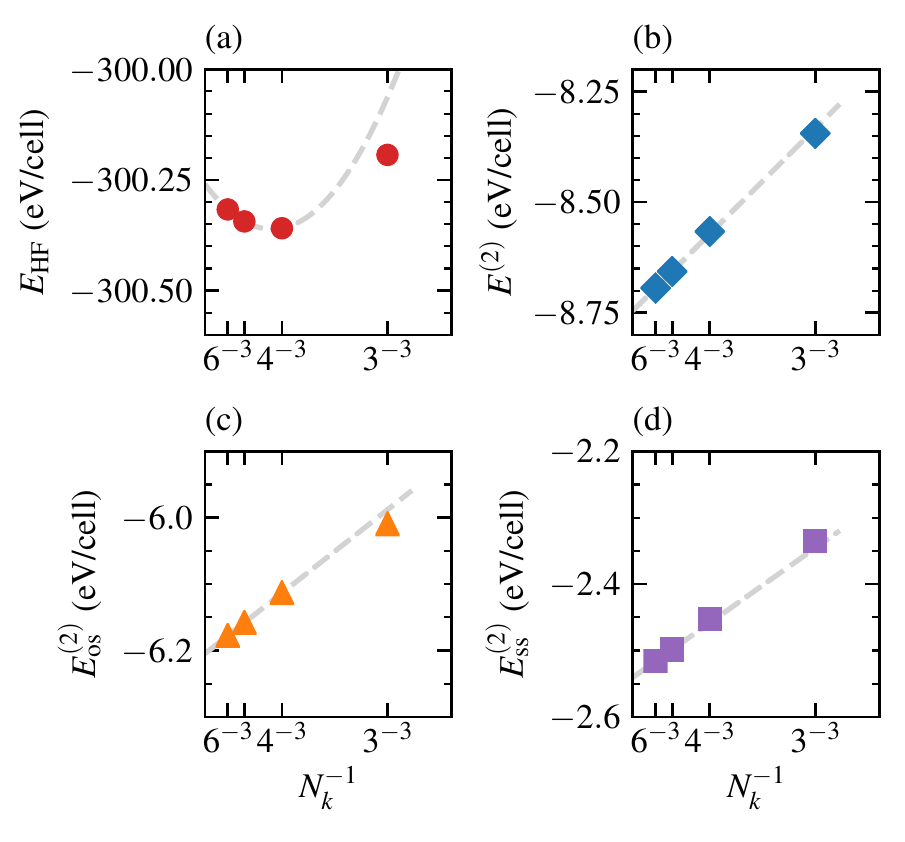}
\caption{Thermodynamic limit convergence of electronic energies for diamond in
the cc-pVTZ basis.  Shown are the HF energy (a), MP2 correlation energy (b), and
the opposite spin (c) and same spin (d) contributions to the MP2 correlation
energy.  Extrapolation to the thermodynamic limit is indicated as a dashed line.
}
\label{fig:c_tdl}
\end{figure}

Due to the compact arrangement of atoms in crystals, linear dependencies are a
common occurrence when using large atom-centered basis sets~\cite{Daga2020,Lee2021,Ye2022}. 
To avoid associated numerical problems, we removed the most diffuse primitive
Gaussians from the orbital basis set of Mg (exponents $<0.05$) and Li 
(exponents $<0.1$), except in LiCl.  From the even-tempered
auxiliary basis set for all systems, we removed diffuse Gaussians with exponents
$<0.2$.  Testing indicated that these modifications do not affect our results.

To assess our ability to reliably access the thermodynamic limit (TDL) and
complete basis set (CBS) limit, we first consider carbon diamond at its
experimental crystal volume.  In Fig.~\ref{fig:c_tdl}, we show convergence to
the TDL using the TZ basis set and $k$-point meshes of $N_k=3^3$--$6^3$ for HF
and $N_k=3^3$--$6^3$ for MP2.  To address the divergent exchange term in
periodic HF, we used a Madelung constant
correction~\cite{Paier2005,Broqvist2009,Sundararaman2013}, which yields total
energies and orbital energies that converge to the TDL as $N_k^{-1}$.
Subsequent MP2 correlation energies converge at the same rate because the
orbital pair densities appearing in the electron repulsion integrals are
chargeless by orthogonality of the molecular orbitals.  
We extrapolate the HF energy according to the form
$E_\mathrm{HF}(N_k) = E_\mathrm{HF}(\infty) + aN_k^{-1} + bN_k^{-2}$
and the MP2 correlation energy according to the form
$E^{(2)}(N_k) = E^{(2)}(\infty) + aN_k^{-1}$
(numerically, we found that the inclusion of sub-leading corrections to the
HF energy was
most appropriate, as shown in Fig.~\ref{fig:c_tdl}, although nearly identical
results are obtained with $b=0$).
We estimate by eye that these extrapolations give results in the TDL that are
accurate to about 0.01~eV. The same-spin and opposite-spin correlation energies
exhibit finite-size errors that are almost identical in magnitude, at least for
this material.

\begin{figure}[t]
\centering
\includegraphics[scale=0.9]{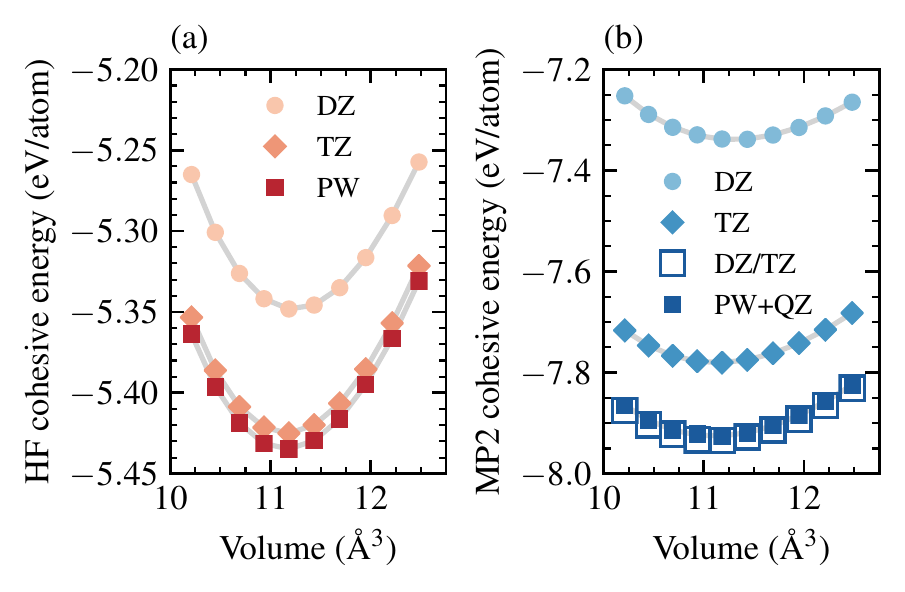}
\caption{
Basis set convergence of the equation of state (EOS) for diamond using a
$4\times4\times4$ $k$-point mesh.  Shown are the HF EOS (a) and MP2 EOS (b),
comparing results obtained with the cc-pVDZ (circles) and cc-pVTZ (diamonds)
basis sets to those obtained using a converged PW basis set for the HF energy
and a PW-resolved cc-pVQZ basis set for the unoccupied orbitals in the
calculation of the MP2 correlation energy (filled squares).  For the MP2 EOS
(b), we also show results obtained using $X^{-3}$ DZ/TZ extrapolation of the MP2
correlation energy (open squares).
}
\label{fig:c_eos}
\end{figure}

In Fig.~\ref{fig:c_eos}, we show the HF (a) and MP2 (b) equation of state (EOS)
of diamond using $N_k=4^3$ and the DZ and TZ basis sets.  For the HF
EOS, we compare to results obtained using a plane-wave (PW) basis set, with a
kinetic energy cutoff chosen to achieve convergence to better than 1 meV/atom.
We see excellent agreement between the PW result and the
TZ result (better than 0.02 eV/atom), indicating that the latter is near the CBS limit. 
For the MP2 energy, which is more expensive to evaluate in a large PW basis, we
compare to a calculation with virtual orbitals obtained from an approximate PW resolution of the QZ
basis set~\cite{Booth2016}, denoted as PW+QZ. 
We see more sensitivity to the basis set, as expected.
Comparing the TZ and PW+QZ results, the lattice constant, bulk modulus, 
and cohesive energy differ by 0.003~\AA, 3.5~GPa, and 0.15~eV
(see below for details about the calculation of these properties).
A $X^{-3}$ CBS extrapolation of the DZ and TZ correlation energies (where $X=2,3$)
gives an EOS in excellent agreement with the PW+QZ result; the deviations are
0.0001~\AA, 1.8~GPa, and 0.009~eV.

To summarize, in all subsequent production calculations, we use the TZ HF energy
and the DZ/TZ CBS-extrapolated MP2 correlation energy and extrapolate to the TDL
using results obtained with $N_k=4^3$--$6^3$ (HF) and $N_k=5^3$--$6^3$ (MP2).

\section{Results and discussion}
\label{sec:results}

We studied 12 semiconductors and insulators, with diamond (C and Si), zincblende
(SiC, BN, BP, AlN, and AlP), and rock salt (MgO, MgS, LiH, LiF, and LiCl)
crystal structures.  These solids were chosen based on their basis set
availability, relatively simple crystal structures, and nonzero bandgap, as
required for the application of MP2-based methods.  We predicted the lattice
constant $a$, bulk modulus $B$, and cohesive energy $E_\mathrm{coh}$ of all
solids using HF, MP2, and SCS/SOS-MP2 by calculating the total energy for ten
different volumes in the range of $\pm10\%$ of the experimental volume. We then
fitted the energies to an equation of state given by a third-order
Birch-Murnaghan form.  The cohesive energy is calculated at the theoretically
predicted equilibrium volume and is defined with respect to the energy of
isolated atoms.  Atomic energies of open-shell atoms were calculated with
unrestricted HF and MP2 and the basis set superposition error was accounted for
by adding crystalline basis functions.

\begin{table}[b]
    \label{tab:summary}
    \centering
    \caption{
        Summary of results for the mean absolute error of the lattice constant
        $a$, bulk modulus $B$, and cohesive energy $E_\mathrm{coh}$, compared to
        experiment.  DFT results (PBE, PBEsol, SCAN) and experimental values, which have
        been corrected for zero-point motion, are from Ref.~\onlinecite{Tran2016}, except for
        those for MgS, which are from Ref.~\onlinecite{Zhang2018}.
    }
    \label{tab:summary}
    \begin{ruledtabular}
    \begin{tabular}{lx{3}dd}
        Method & \multicolumn{1}{c}{$a$ (\AA)} & 
                 \multicolumn{1}{c}{$B$ (GPa)} & 
                 \multicolumn{1}{c}{$E_\mathrm{coh}$ (eV)} \\
        \hline
        HF [ref~\onlinecite{Gruneis2010}] & 0.059 & 12.3 & 1.59 \\
        HF & 0.057 & 13.5 & 1.54 \\
        PBE & 0.061 & 12.2 & 0.19 \\
        PBEsol & 0.030 & 7.8 & 0.31 \\
        SCAN & 0.030 & 7.4 & 0.19 \\
        MP2 [ref~\onlinecite{Gruneis2010}] & 0.021 & 6.4 & 0.23 \\
        MP2 & 0.020 & 5.2 & 0.23 \\
        SCS-MP2 & 0.015 & 3.8 & 0.06 \\
        SOS-MP2 & 0.017 & 3.7 & 0.08 \\
    \end{tabular}
    \end{ruledtabular}
\end{table}

As an initial test of our methods and implementation, we have compared our
calculated HF and MP2 results to those of Gr\"uneis et al.~\cite{Gruneis2010},
who studied 11 of the 12 solids studied here at the same levels of theory (all
except MgS).  At the HF level, the mean absolute deviations in the lattice
parameter, bulk modulus, and cohesive energy are $0.010~\AA$, 2.2~GPa, and
0.09~eV.  At the MP2 level, the same deviations are $0.017~\AA$, 3.6~GPa, and
0.05~eV.  A few of the biggest discrepancies are for the lattice constant of LiF
($0.039~\AA$ with HF and $0.036~\AA$ with MP2); the bulk moduli of diamond
(5.6~GPa with HF and 16.8~GPa with MP2), SiC (5.1~GPa with MP2), and MgO (5 GPa
with HF and MP2); and the cohesive energies of SiC (0.37~eV with HF) and AlP
(0.25~eV with HF).  Nonetheless, these mean absolute deviations provide a rough
estimate of the precision of our HF and MP2 calculations due to pseudopotential,
basis set, and finite-size errors, complementing the detailed study of carbon
diamond in the previous section.  Agreement with experiment to higher accuracy
should be viewed with caution. 

For most materials, we will compare to the experimental values compiled in
Ref.~\onlinecite{Zhang2018}, from which we remove zero-point energy
contributions calculated at the PBE level in the same work.  For AlN and LiH,
which are absent in Ref.~\onlinecite{Zhang2018}, we use values from
Ref.~\onlinecite{Schimka2011}, which were identically corrected for zero-point
energy in that work.  For most of the shared materials, these two references
give very similar properties; three large outliers are the bulk moduli of BN
(differing by 21.7 GPa), BP (8.5 GPa), and MgO (3.2 GPa).

\begin{figure}[b]
\centering
\includegraphics[scale=0.9]{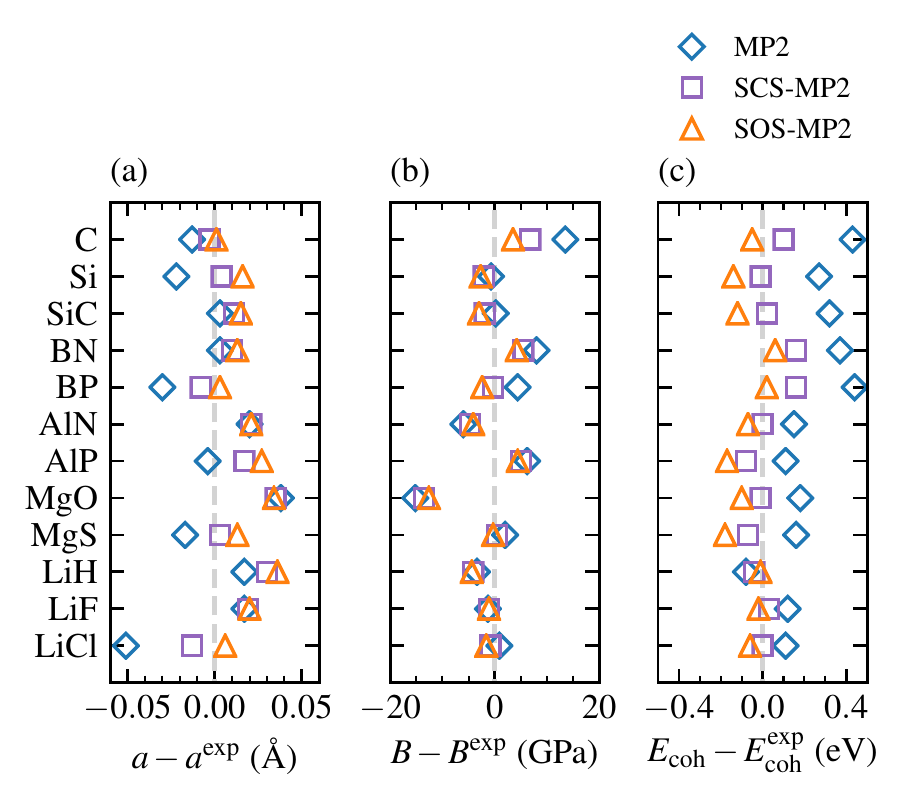}
\caption{ 
Errors in the lattice constant $a$ (a), bulk modulus $B$ (b), and
cohesive energy $E_\mathrm{coh}$ (c) compared to experimental values.
}
\label{fig:ae_mol}
\end{figure}

\begin{figure*}[t]
\begin{center}
\includegraphics[scale=0.9]{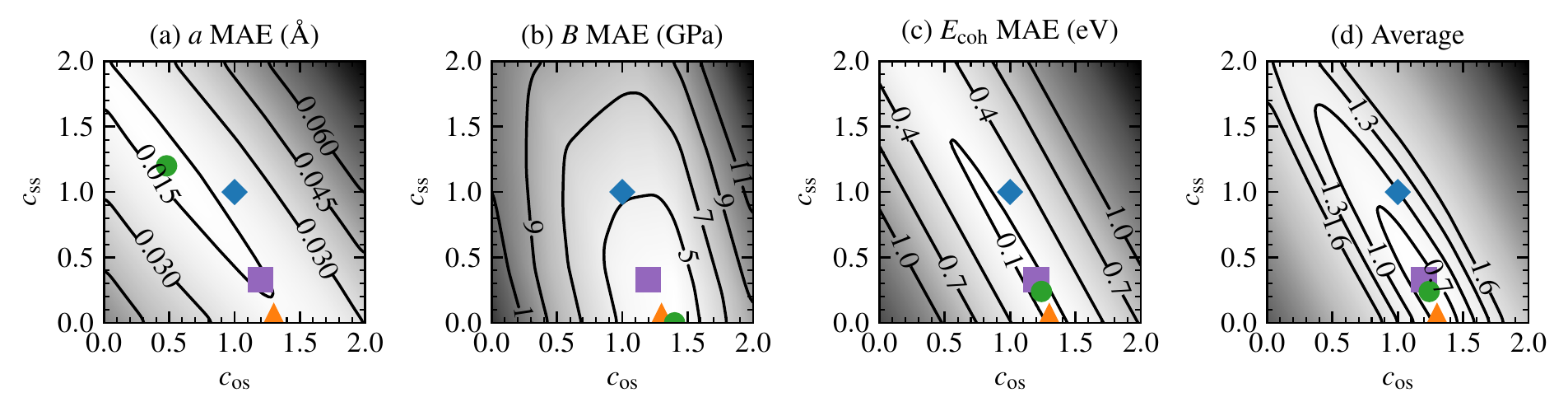}
\end{center}
\caption{
MAE for the SCS-MP2 lattice constant (a), bulk modulus (b), cohesive
energy (c), and dimensionless average (d) calculated according to
Eq.~(\ref{eq:avg}).  Indicated points correspond to the scaling parameters for
conventional MP2 (blue diamond), SCS-MP2 (purple square), SOS-MP2 (orange
triangle), and property-specific optimal SCS-MP2 (green circle).
}
\label{fig:mae_tot}
\end{figure*}

A summary of our results is given in Tab.~\ref{tab:summary}, which shows the
mean absolute error (MAE) of various methods for each property compared to these
experimental values; material-specific predictions for all properties are given
in Tabs.~\ref{table:a}, \ref{table:b}, and \ref{table:e} in App.~\ref{app:tabs}.
HF predictions are unsurprisingly poor.
MP2 predictions are a significant improvement, exhibiting MAEs of 0.020~$\AA$,
5.2~GPa, and 0.23~eV.  SCS-MP2 predictions are a further improvement, exhibiting
MAEs of 0.015~$\AA$, 3.8~GPa, and 0.06~eV; the cohesive energy is especially
improved.  The SOS-MP2 predictions are similarly good, exhibiting MAEs of
0.017~$\AA$, 3.7~GPa, and 0.08~eV.  We therefore conclude that SCS-MP2 and
SOS-MP2, using the most common spin-scaling parameters, perform remarkably well
for the prediction of the structural and energetic properties of
three-dimensional covalent and ionic semiconductors and insulators.

The performance of MP2-based methods can be broken down for each material, as
shown in Fig.~\ref{fig:ae_mol}. Some of the apparent outliers at the MP2 level
of theory are the lattice constant of LiCl, MgO and BP; the bulk modulus of C
and MgO; and the cohesive energy of C, BN, and BP. Generally, we see that spin
scaling yields an increase in the lattice constant, a decrease in the bulk
modulus, and a decrease in the cohesive energy.  The latter, in particular,
corrects the tendency of MP2 to overestimate the cohesive energy, as it does for
all materials except LiH.

We finally assess to what extent further improvement is possible by a
reoptimization of the spin-scaling parameters.  In Fig.~\ref{fig:mae_tot} we
plot the SCS-MP2 MAE with respect to experiment of the lattice constant, bulk
modulus, and cohesive energy as a function of the spin scaling parameters.  In
addition to the MP2 and molecular SCS-MP2 parameters, we indicate the optimal
SCS and SOS parameters for each property, and we see that different properties
favor different combinations of the spin scaling coefficients.  Moreover, many
of the landscape valleys are long and narrow, indicating a wide range of
parameters that deliver comparable performance for a given property. To identify
a globally optimal set of parameters, we calculate a dimensionless average
\begin{equation}
\label{eq:avg}
\mathrm{Avg} = \frac{1}{3}\left(
    \frac{\langle|\Delta a|\rangle}{\langle|\Delta a|\rangle_{\mathrm{MP2}}}
    +\frac{\langle|\Delta B|\rangle}{\langle|\Delta B|\rangle_{\mathrm{MP2}}}
    +\frac{\langle |\Delta E_\mathrm{coh}|\rangle}{\langle|\Delta E_\mathrm{coh}|\rangle_{\mathrm{MP2}}}\right)
\end{equation}
where $\langle|\Delta O|\rangle$ indicates the MAE of property $O$
and $\langle \cdot \rangle_\mathrm{MP2}$ indicates the MP2 value.  The
weighting is such that the value of the cost function is the average error with
respect to that of MP2; of course, different ways of averaging will give
different optimal parameters.  This average error is plotted in
Fig.~\ref{fig:mae_tot}(d) and identifies the optimal SCS parameters
$(c_\mathrm{os}, c_\mathrm{ss})=(1.24,0.24)$ and optimal SOS parameter
$c_\mathrm{os}=1.36$, which are quite similar to the standard values determined
for molecules.  The optimal parameters of the average are very similar to those
of the cohesive energy, in part because the latter's optimal parameters balance
the errors of the lattice constant and the bulk modulus.  Moreover, the average
error of the standard SCS and SOS parameters is only marginally higher: the
optimal values give 0.568 (SCS) and 0.585 (SOS) while the standard values give
and 0.570 (SCS) and 0.646 (SOS).  In other words, spin-component scaling in any
of these nearly optimal forms yields MAEs that are about 60\% those of MP2.

\section{Conclusion}
\label{sec:conclusion}

We have studied the performance of periodic SCS- and SOS-MP2 in the TDL and CBS
limit, finding excellent agreement with experimental values. As shown in
Tab.~\ref{tab:summary} the performance is significantly better than popular
functionals for solid-state calculations, including
PBE~\cite{perdew1996generalized}, PBEsol~\cite{Perdew2008}, and
SCAN~\cite{Sun2015}.  The performance of SOS-MP2 is only marginally worse than
SCS-MP2, indicating excellent promise as an especially affordable technique.
Although we have carefully addressed finite-size and basis set errors, there may
be residual pseudopotential errors, which are harder to remove systematically.
For example, preliminary testing indicates that MP2 calculations on MgO, which
here exhibited atypically large errors in the lattice constant and bulk modulus,
are significantly improved through the use of a small-core pseudopotential with
a [He] core.  A careful study of pseudopotentials and core-correlation effects
is in progress and will be presented elsewhere~\cite{YeECP}.

The conclusions of this work are limited to the class of materials studied,
i.e.~three-dimensional covalent and ionic insulators, and future work must
assess the extension to surfaces, layered materials, and other weakly-bound
solids, such as molecular crystals.  An early periodic MP2 study by Del Ben et
al.~\cite{DelBen2012} observed good SCS-MP2 and double hybrid performance for
molecular crystals, and a recent report from our group~\cite{Bintrim2022} found
that SCS-MP2 yields a very accurate prediction of the cohesive energy of the
benzene crystal, while SOS-MP2 yields a nonnegligible underestimation.  The
present work has demonstrated good performance over a relatively wide range of
spin scaling parameters, indicating that additional systems or properties can be
incorporated in the identification of optimal parameters for condensed phase
materials.  Perhaps most importantly, we have found that the standard scaling
parameters, which are known to deliver good performance for molecular chemistry,
are almost optimal for solid-state properties. This transferability suggests
that accurate surface chemistry on insulators should be accurately described by
SCS- and SOS-MP2.

\section*{Acknowledgments}

This work was supported by the National Science Foundation under Grant No.\
CHE-1848369 (T.G.) and Grant No.~OAC-1931321 (H.-Z.Y.).  
We acknowledge computing resources from Columbia University's
Shared Research Computing Facility project, which is supported by NIH Research
Facility Improvement Grant 1G20RR030893-01, and associated funds from the New
York State Empire State Development, Division of Science Technology and
Innovation (NYSTAR) Contract C090171, both awarded April 15, 2010. The Flatiron
Institute is a division of the Simons Foundation.

\section*{Data availability statement}
The data that support the findings of this study are available from the
corresponding author upon reasonable request.

\appendix
\section{Material-specific predicted properties}
\label{app:tabs}

In Tabs.~\ref{table:a}, \ref{table:b}, and \ref{table:e}, we provide the material-specific
predictions at the HF, MP2, SCS-MP2, and SOS-MP2 levels of theory, along with
their mean absolute error (MAE) and mean absolute relative error (MARE) compared
to experimental values.

\begin{table}
\begin{ruledtabular}
\begin{tabular}{l c c c c c}
\multicolumn{6}{c}{Lattice constant ($\AA$)} \\
Solid & HF & MP2 & SCS-MP2 & SOS-MP2 & Exp. \\
\hline
   C & 3.547 & 3.540 & 3.550 & 3.554 & 3.553 \\
  Si & 5.508 & 5.399 & 5.425 & 5.437 & 5.421 \\
 SiC & 4.371 & 4.350 & 4.358 & 4.362 & 4.347 \\
  BN & 3.596 & 3.596 & 3.603 & 3.606 & 3.593 \\
  BP & 4.584 & 4.495 & 4.517 & 4.528 & 4.525 \\
 AlN & 4.365 & 4.388 & 4.389 & 4.389 & 4.368 \\
 AlP & 5.542 & 5.444 & 5.465 & 5.475 & 5.448 \\
 MgO & 4.176 & 4.227 & 4.224 & 4.223 & 4.189 \\
 MgS & 5.281 & 5.171 & 5.191 & 5.201 & 5.188 \\
 LiH & 4.094 & 3.996 & 4.009 & 4.015 & 3.979 \\
 LiF & 3.964 & 3.990 & 3.992 & 3.993 & 3.973 \\
LiCl & 5.253 & 5.021 & 5.059 & 5.078 & 5.072 \\
\hline
MAE (\AA) & 0.057 & 0.020 & 0.015 & 0.017 & -- \\
MARE (\%) &   1.1 &   0.4 &   0.3 &   0.3 & -- \\

\end{tabular}
\end{ruledtabular}
\caption{Predicted lattice constants at the indicated level of theory, including
mean absolute error (MAE) and mean absolute relative error (MARE). Experimental
values, which have been corrected for zero-point motion, are from Ref.~\onlinecite{Zhang2018},
except for those for AlN and LiH, which are from Ref.~\onlinecite{Schimka2011}.}
\label{table:a}
\end{table}

\begin{table}
\begin{ruledtabular}
\begin{tabular}{l c c c c c}
\multicolumn{6}{c}{Bulk modulus (GPa)} \\
Solid & HF & MP2 & SCS-MP2 & SOS-MP2 & Exp. \\
\hline
   C & 500.6 & 466.8 & 460.1 & 456.8 & 453.3 \\
  Si & 102.1 &  99.6 &  98.2 &  97.6 & 100.3 \\
 SiC & 242.7 & 229.1 & 227.0 & 225.9 & 228.9 \\
  BN & 430.2 & 396.5 & 394.0 & 392.7 & 388.5~\cite{Zhang2018}, 410.2~\cite{Schimka2011} \\
  BP & 174.8 & 180.9 & 176.2 & 174.1 & 176.5~\cite{Zhang2018}, 168.0~\cite{Schimka2011} \\
 AlN & 226.1 & 200.0 & 201.3 & 201.9 & 206.0 \\
 AlP &  93.9 &  93.2 &  92.0 &  91.4 &  87.0 \\
 MgO & 180.2 & 157.8 & 159.5 & 160.4 & 173.0~\cite{Zhang2018}, 169.8~\cite{Schimka2011} \\
 MgS &  76.9 &  83.0 &  81.4 &  80.7 &  81.0 \\
 LiH &  32.1 &  36.7 &  36.0 &  35.7 &  40.1 \\
 LiF &  77.4 &  74.1 &  74.3 &  74.3 &  75.4 \\
LiCl &  29.6 &  38.2 &  36.5 &  35.7 &  37.3 \\
\hline
MAE (GPa) & 13.5 &  5.2 &  3.8 &  3.7 & -- \\
MARE (\%) & 36.3 & 13.8 & 10.3 &  9.9 & -- \\

\end{tabular}
\end{ruledtabular}
\caption{Predicted bulk moduli at the indicated level of theory, including
mean absolute error (MAE) and mean absolute relative error (MARE). Experimental
values, which have been corrected for zero-point motion, are from Ref.~\onlinecite{Zhang2018},
except for those for AlN and LiH, which are from Ref.~\onlinecite{Schimka2011}.
For three large discrepancies (BN, BP, MgO), we include results from both references.}
\label{table:b}
\end{table}

\begin{table}
\begin{ruledtabular}
\begin{tabular}{l c c c c c}
\multicolumn{6}{c}{Cohesive energy (eV)} \\
Solid & HF & MP2 & SCS-MP2 & SOS-MP2 & Exp. \\
\hline
   C & 5.38 & 7.98 & 7.65 & 7.50 & 7.55 \\
  Si & 3.03 & 4.97 & 4.69 & 4.56 & 4.70 \\
 SiC & 4.53 & 6.79 & 6.49 & 6.35 & 6.47 \\
  BN & 4.78 & 7.13 & 6.92 & 6.82 & 6.76 \\
  BP & 3.42 & 5.58 & 5.30 & 5.16 & 5.14 \\
 AlN & 3.86 & 6.00 & 5.85 & 5.78 & 5.85 \\
 AlP & 2.71 & 4.42 & 4.23 & 4.14 & 4.31 \\
 MgO & 3.62 & 5.37 & 5.18 & 5.09 & 5.19 \\
 MgS & 2.78 & 4.20 & 3.97 & 3.86 & 4.04 \\
 LiH & 1.85 & 2.41 & 2.45 & 2.48 & 2.49 \\
 LiF & 3.41 & 4.58 & 4.49 & 4.44 & 4.46 \\
LiCl & 2.73 & 3.69 & 3.58 & 3.52 & 3.58 \\
\hline
MAE (eV)  & 1.54 & 0.23 & 0.06 & 0.08 & -- \\
MARE (\%) & 43.0 &  6.4 &  1.6 &  2.4 & -- \\

\end{tabular}
\end{ruledtabular}
\caption{Predicted cohesive energies at the indicated level of theory, including
mean absolute error (MAE) and mean absolute relative error (MARE). Experimental
values, which have been corrected for zero-point motion, are from Ref.~\onlinecite{Zhang2018},
except for those for AlN and LiH, which are from Ref.~\onlinecite{Schimka2011}.}
\label{table:e}
\end{table}

\end{document}